\documentclass[conference]{IEEEtran}
\IEEEoverridecommandlockouts
\usepackage{cite}
\usepackage{amsmath,amssymb,amsfonts}
\usepackage{algorithmic}
\usepackage{graphicx}
\usepackage{textcomp}
\usepackage{xcolor}
\usepackage{hyperref}
\usepackage{booktabs}
\usepackage{multirow}

\hypersetup{
    colorlinks=true, 
    linkcolor=black,  
    citecolor=black, 
    filecolor=black, 
    urlcolor=black    
}

\def\BibTeX{{\rm B\kern-.05em{\sc i\kern-.025em b}\kern-.08em
    T\kern-.1667em\lower.7ex\hbox{E}\kern-.125emX}}
\begin{document}

\title{Consensus-based Distributed Quantum Kernel Learning for Speech Recognition
\thanks{*The first two authors contributed equally to this work.\\}
}

\author{
\IEEEauthorblockN{
    Kuan-Cheng Chen\IEEEauthorrefmark{2}\IEEEauthorrefmark{3}\IEEEauthorrefmark{4}\IEEEauthorrefmark{1}, 
    Wenxuan Ma\IEEEauthorrefmark{5}\IEEEauthorrefmark{1}
    Xiaotian Xu\IEEEauthorrefmark{3}
}
\IEEEauthorblockA{\IEEEauthorrefmark{2}Department of Electrical and Electronic Engineering, Imperial College London, London, UK}
\IEEEauthorblockA{\IEEEauthorrefmark{3}Centre for Quantum Engineering, Science and Technology (QuEST), Imperial College London, London, UK}
\IEEEauthorblockA{\IEEEauthorrefmark{5}College of Control Science and Engineering, Zhejiang University, Hangzhou, China}
\IEEEauthorblockA{Email: 
\IEEEauthorrefmark{4}kuan-cheng.chen17@imperial.ac.uk
}
}

\maketitle

\begin{abstract}
This paper presents a Consensus-based Distributed Quantum Kernel Learning (CDQKL) framework aimed at improving speech recognition through distributed quantum computing.CDQKL addresses the challenges of scalability and data privacy in centralized quantum kernel learning. It does this by distributing computational tasks across quantum terminals, which are connected through classical channels. This approach enables the exchange of model parameters without sharing local training data, thereby maintaining data privacy and enhancing computational efficiency. Experimental evaluations on benchmark speech emotion recognition datasets demonstrate that CDQKL achieves competitive classification accuracy and scalability compared to centralized and local quantum kernel learning models. The distributed nature of CDQKL offers advantages in privacy preservation and computational efficiency, making it suitable for data-sensitive fields such as telecommunications, automotive, and finance. The findings suggest that CDQKL can effectively leverage distributed quantum computing for large-scale machine-learning tasks.
\end{abstract}


\section{Introduction}

Quantum computing harnesses the principles of quantum mechanics, such as entanglement and superposition, to tackle complex computational problems, often achieving exponential speedups that classical algorithms cannot match\cite{yamasaki2020learning}. For example, variational quantum algorithms (VQAs)\cite{cerezo2021variational, liu2023learning}, which integrate quantum computing with optimization and machine learning, have become a leading approach for demonstrating quantum advantage. As classical machine learning models struggle with the growing complexity and scale of data, quantum computing presents a promising pathway to address these challenges\cite{huang2021power}. The formal distinction between classical and quantum learnability underscores the transformative potential of quantum computing paradigms in revolutionizing machine learning and computational sciences\cite{jerbi2024shadows}.

\begin{figure}[!t]
    \centering
    \includegraphics[width=\linewidth]{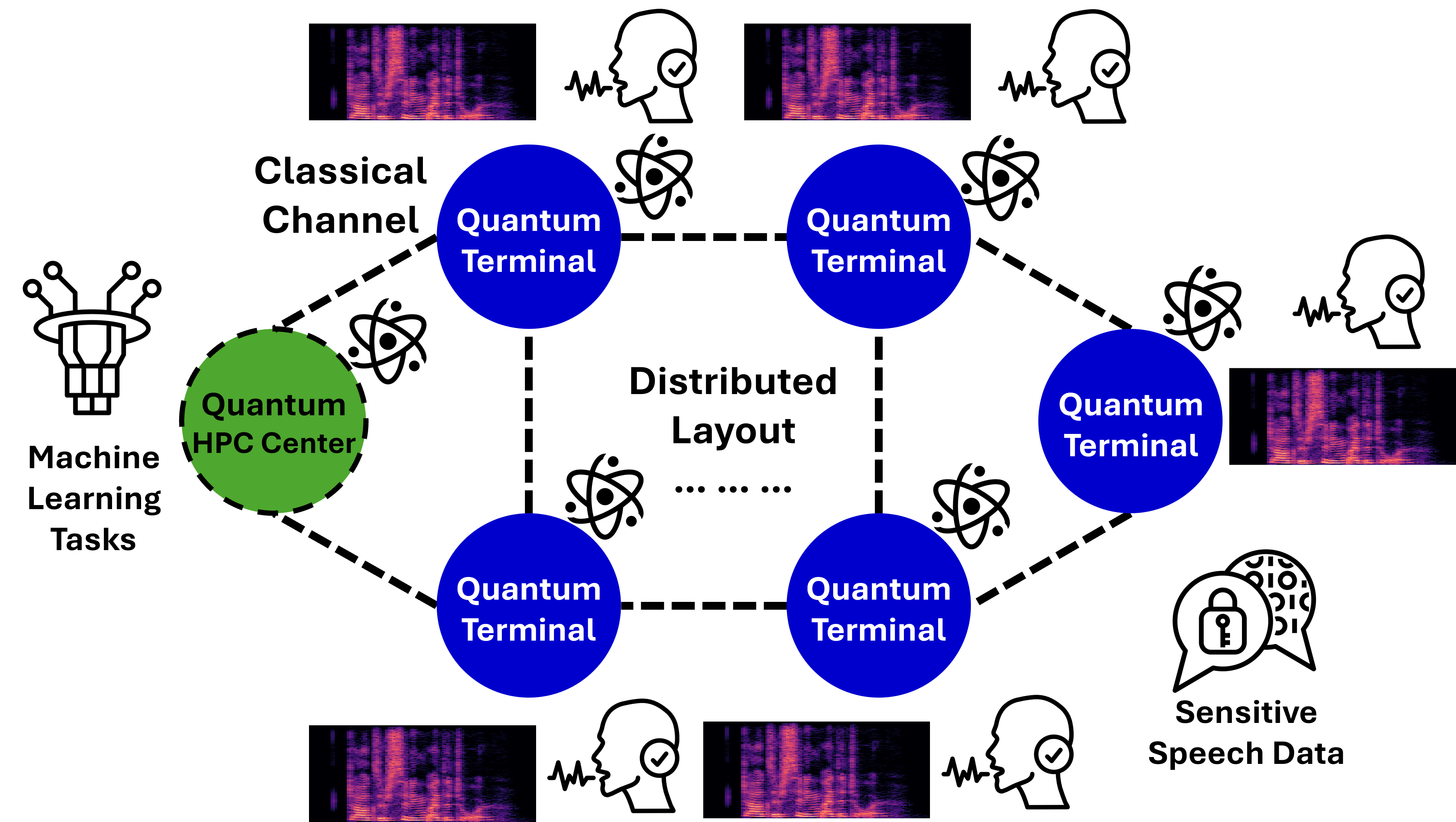}
    \caption{Conceptual Framework of CDQKL for Speech Recognition in a Quantum HPC Distributed System. The diagram shows quantum terminals in a distributed network processing sensitive speech data locally, preserving privacy while exchanging model parameters via classical channels, coordinated by a central Quantum HPC center, to enhance scalability and computational efficiency.}
    \label{fig:conceptual}
\end{figure}

Kernel learning, although not at the forefront of contemporary machine learning research, remains a valuable framework for addressing nonlinear problems by transforming data into higher-dimensional spaces \cite{cervantes2020comprehensive}. Quantum Kernel Learning (QKL) extends this classical approach into the quantum domain, leveraging the unique properties of quantum mechanics to enhance computational speed and representational power by mapping features into the quantum state Hilbert space \cite{schuld2019quantum, rebentrost2014quantum}. The effectiveness of quantum support vector machines with a large number of qubits has been demonstrated using tensor network-based quantum circuit simulations (cuTN-QSVM) \cite{chen2024cutn}. Recent studies have further highlighted the potential of QKL in tackling high-dimensional, large-scale machine learning tasks through practical applications and demonstrations \cite{chen2024quantum, ho2024quantum}. Despite these potential advantages in computational efficiency and expressiveness, QKL faces significant scalability challenges due to the high computational costs associated with constructing large quantum kernel matrices, which become increasingly prohibitive as dataset sizes grow. Moreover, centralized quantum kernel learning struggles with managing sensitive data, such as medical, personal, and financial information, raising critical concerns about data privacy and security.

To address these challenges, this paper introduces to apply a novel consensus-based distributed quantum kernel learning (termed CDQKL \cite{ma2024cdqkl}) approach to the senario of speech recognition. This approach tailored to enhance scalability and preserve data privacy in speech recognition. CDQKL draws inspiration from federated learning and distributed quantum computing by enabling model parameter exchange between adjacent nodes without sharing local training data, thus maintaining data privacy and security\cite{ma2024cdqkl}. As shown in Fig. \ref{fig:conceptual}, this distributed architecture involves a network of quantum terminals connected through classical channels, allowing for large-scale machine learning tasks without direct data aggregation. As quantum devices are increasingly deployed in cloud services and localized environments, CDQKL aligns with this evolving landscape, offering a compelling solution that leverages distributed data processing to improve training efficiency and scalability in complex machine learning tasks, particularly when handling sensitive data.

The distributed nature of CDQKL offers significant advantages for privacy preservation, making it particularly impactful in industries that handle sensitive information, such as telecommunications, automotive, finance, and healthcare. Applications include speech and facial recognition, autonomous driving systems, and financial analytics, where data security and scalability are critical. Unlike previous research in distributed quantum computing \cite{burt2024generalised, chen2024noise}, the proposed algorithm processes sensitive data locally at each quantum terminal, exchanging only model parameters between nodes. This approach enhances training efficiency while mitigating the privacy risks associated with centralized data processing. This paper pioneers the application of CDQKL to speech signal processing, demonstrating its effectiveness through extensive experiments in terms of accuracy, convergence speed, and scalability compared to centralized and local QKL approaches, as well as traditional kernel classifiers. The algorithm’s robust performance underscores its potential to transform data-driven applications in sectors that demand high computational efficiency and stringent privacy standards.

\section{Methodology}

\subsection{Classical Kernel Learning and Quantum Kernel Learning}
SVMs effectively handle high-dimensional, non-linear classification problems by identifying an optimal hyperplane in an \( N \)-dimensional space with feature vectors \( \mathbf{x}_i \in \mathbb{R}^N \) and labels \( y_i \). The goal is to maximize the margin between classes, enhancing generalization to new data. The optimization problem is expressed as:

\begin{equation}
\min_{\mathbf{w}_k, b_k, \zeta_k} \frac{1}{2} \|\mathbf{w}_k\|^2 + C \sum_{i=1}^M \zeta_{ik},
\end{equation}

\begin{equation}
\text{subject to } y_{ik}(\mathbf{w}_k \cdot \mathbf{x}_i + b_k) \geq 1 - \zeta_{ik}, \quad \zeta_{ik} \geq 0,
\end{equation}

where \( \mathbf{w}_k \), \( b_k \), and \( \zeta_{ik} \) represent hyperplane parameters and slack variables. Kernel methods extend SVMs to non-linear spaces using the kernel function:

\begin{equation}
K(\mathbf{x}_i, \mathbf{x}_j) = \phi(\mathbf{x}_i) \cdot \phi(\mathbf{x}_j),
\end{equation}

where \( \phi \) maps data into a higher-dimensional space, capturing complex relationships.

Quantum SVMs (QSVMs) extend classical SVMs into the quantum domain by mapping data into quantum states within the Hilbert space\cite{rebentrost2014quantum,li2015experimental}. The mapping function \( \phi(\mathbf{x}) \) encodes classical data into quantum states \( |\Phi(\mathbf{x})\rangle \). Rotational entanglement operations, defined as:

\begin{equation}
U_{\Phi}(\mathbf{x}) = \exp\left(i \sum_{S \subseteq [n]} \phi_S(\mathbf{x}) \prod_{k \in S} Z_k \right),
\end{equation}

facilitate the encoding process. The quantum kernel function between two points is calculated as:

\begin{align}
K(\mathbf{x}_i, \mathbf{x}_j) &= |\langle \psi(\mathbf{x}_i) | \psi(\mathbf{x}_j) \rangle|^2 \\
&= |\langle 0^{\otimes N} | U^\dagger(\mathbf{x}_i) U(\mathbf{x}_j) | 0^{\otimes N} \rangle|^2.
\end{align}

leveraging quantum parallelism for computational speedups by \( O(\log(n)) \) for computing the kernel function for \( n^2 \) pairs of data points over classical methods \cite{gentinetta2024complexity, zhang2023quantum}.

\subsection{Consensus-based Distributed Quantum Kernel Learning}

\begin{figure}[!b]
    \centering
    \includegraphics[width=\linewidth]{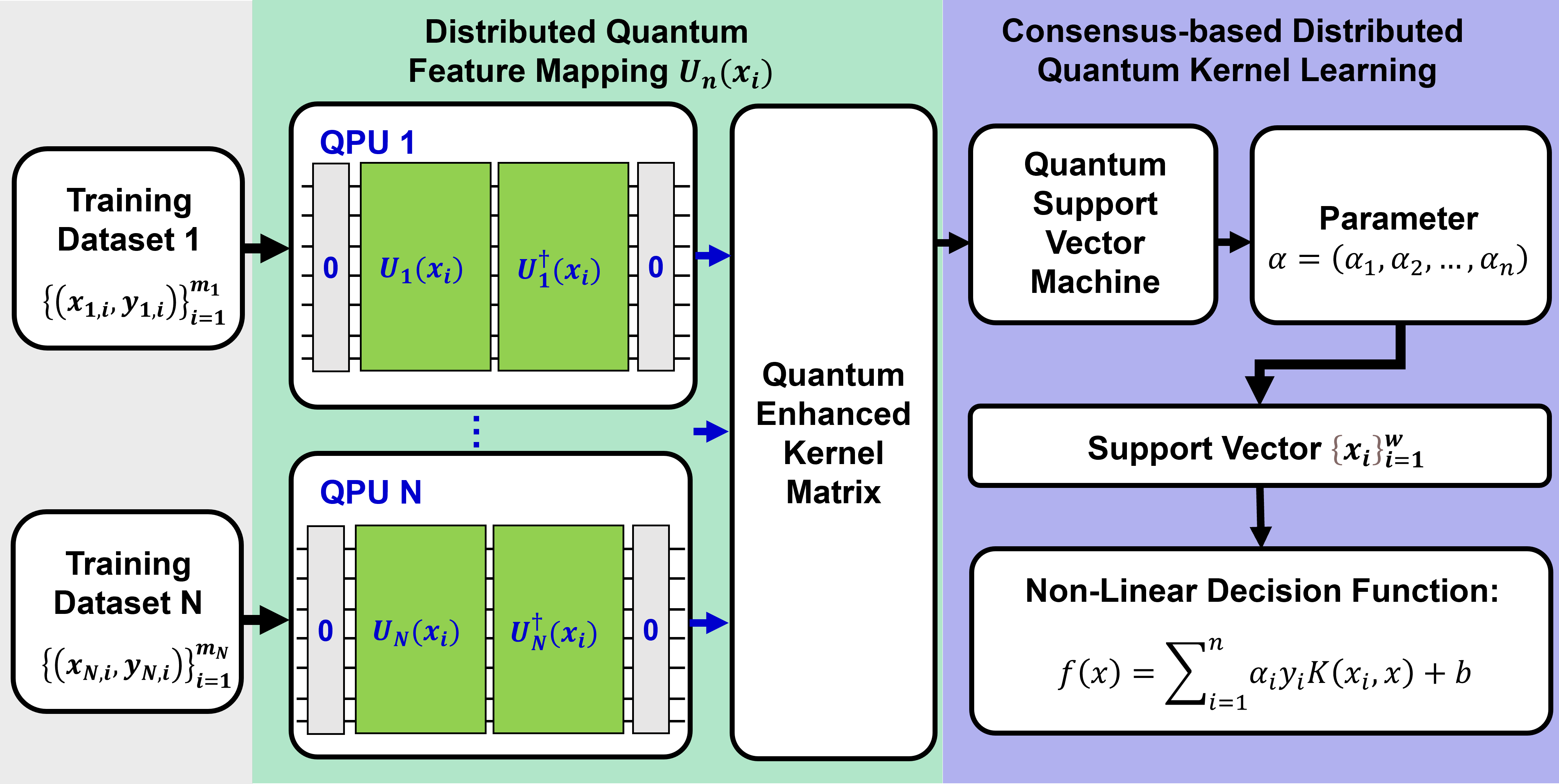}
    \caption{Schematic of the CDQKL framework showing distributed quantum feature mapping across QPUs and the consensus-based learning process for enhancing classification using a Quantum Support Vector Machine.}
    \label{fig:flow}
\end{figure}

In this paper, we utilize a consensus-based distributed quantum kernel learning (CDQKL) framework \cite{ma2024cdqkl} designed for large-scale speech recognition task. Our model consists of \( N \) quantum computing units interconnected through classical communication channels, forming a network represented by a graph \( G(V, E) \), where \( V = \{1, \ldots, N\} \) represents the quantum units, and \( E \subseteq V \times V \) indicates the communication links. For each link \((i, j) \in E\), unit \( j \) is considered a neighbor of unit \( i \), and the neighborhood set of unit \( i \) is \( N_i = \{ j \mid (i, j) \in E, \forall j \in V \} \cup \{i\} \). The communication between nodes is governed by a consensus matrix \( W = [w_{ij}]_{N \times N} \), where \( w_{ij} > 0 \) if \((i, j) \in E\) and \( w_{ij} = 0 \) otherwise.

We consider \( G \) as a connected graph with a doubly stochastic matrix \( W \) that satisfies \( W\mathbf{1}_N = \mathbf{1}_N \) and \( \mathbf{1}_N^T W = \mathbf{1}_N^T \), where \(\mathbf{1}_N\) is an \( N \)-dimensional vector of ones. The properties of doubly stochastic matrices, such as having a spectral radius of 1, provide stability and convergence benefits critical for distributed optimization algorithms.

Each quantum unit \( i \) possesses a local dataset \( D_i \). The objective is to minimize the global loss function defined over the entire network:

\begin{equation}
\min_{\theta \in \mathbb{R}^n} L(D, \theta) = \sum_{i=1}^{N} L_i(D_i, \theta) = -\sum_{i=1}^{N} V_i(K_i(\theta), K^*_i).
\end{equation}

Here, \( D \) represents the entire dataset. \(V_i\) and \( K_i(\theta) \) are alignment values and the quantum kernel associated with the \( i \)-th quantum unit, respectively. This formulation enables distributed quantum kernel learning, where each unit minimizes its local loss function \( L_i(D_i, \theta) \), contributing to the global optimization task.

To achieve consensus across the network, we employ a gradient-based distributed optimization algorithm. At each iteration \( k \), the quantum unit \( i \) updates an auxiliary variable \( \lambda_i^k \) and its parameter vector \( \theta_i^k \) as follows:

\begin{equation}
\lambda_i^k = \sum_{j \in N_i} w_{ij} \theta_j^{k-1}, \quad \theta_i^k = \lambda_i^k - \eta_i \nabla_\theta L_i(D_i, \lambda_i^k),
\end{equation}

where \( \eta_i \) is the step size, and \( \lambda_i^k \) aggregates gradient information from neighboring nodes, facilitating the consensus mechanism. The gradient of the local loss function \( \nabla_\theta L_i(D_i, \lambda_i^k) \) is estimated using \( m_i \) data points \( s_i^p \) from the local dataset:

\begin{equation}
\nabla_\theta L_i(D_i, \lambda_i^k) = \frac{1}{m_i} \sum_{p=1}^{m_i} \nabla_\theta L_p(s_i^p, \lambda_i^k).
\end{equation}

To further enhance computational efficiency, a stochastic gradient descent approach is employed, wherein the gradient is approximated using a random subset of \( q_i \) samples:

\begin{equation}
\tilde{\nabla}_\theta L_i(D_i, \lambda_i^k) = \frac{1}{q_i} \sum_{p=1}^{q_i} \nabla_\theta L_p(s_i^p, \lambda_i^k).
\end{equation}

Thus, each quantum unit \( i \) updates its parameters using the consensus-based stochastic gradient algorithm:

\begin{equation}
\lambda_i^k = \sum_{j \in N_i} w_{ij} \theta_j^{k-1}, \quad \theta_i^k = \lambda_i^k - \eta_i \tilde{\nabla}_\theta L_i(D_i, \lambda_i^k).
\end{equation}

This iterative procedure ensures that the units reach a consensus on the model parameters without the need to share sensitive local data, offering both computational efficiency and data security. The proposed CDQKL framework (shown in Fig. \ref{fig:flow}) demonstrates significant potential for enhancing the scalability and privacy of distributed quantum kernel learning, making it a promising approach for applications in automotive, finance, and other data-sensitive domains.

\section{Experiments and Result Analysis}

\subsection{Experimental Setup}

\begin{figure}[!t]
    \centering
    \includegraphics[width=0.9\linewidth]{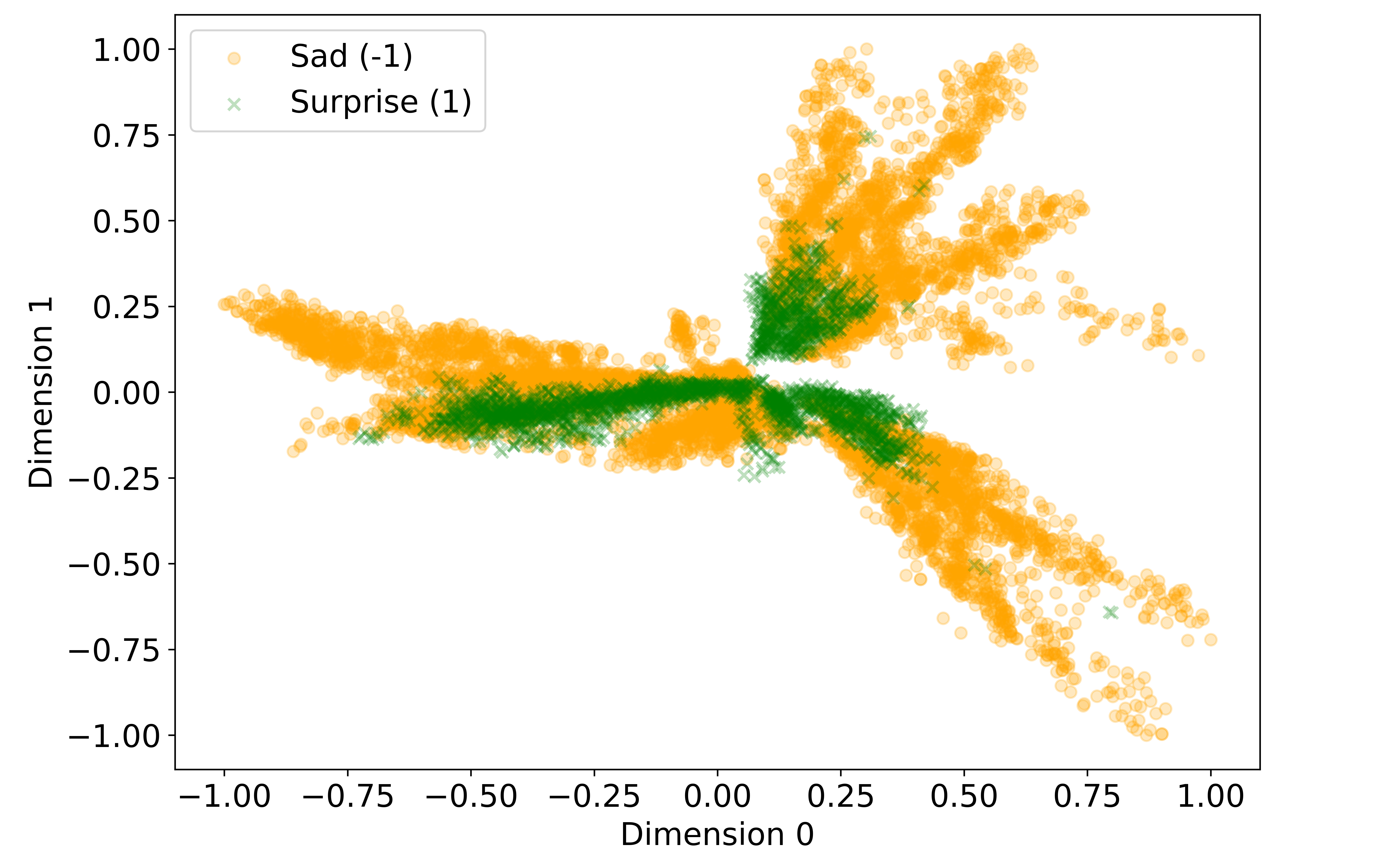}
    \caption{2D projection of training and testing data points for speech emotion recognition, illustrating the challenging separability between ``Sad" (-1) and ``Surprise" (1) emotions in the feature space.}
    \label{fig:sad-sup-dist}
\end{figure}

\begin{figure}[!b]
    \centering
    \includegraphics[width=\linewidth]{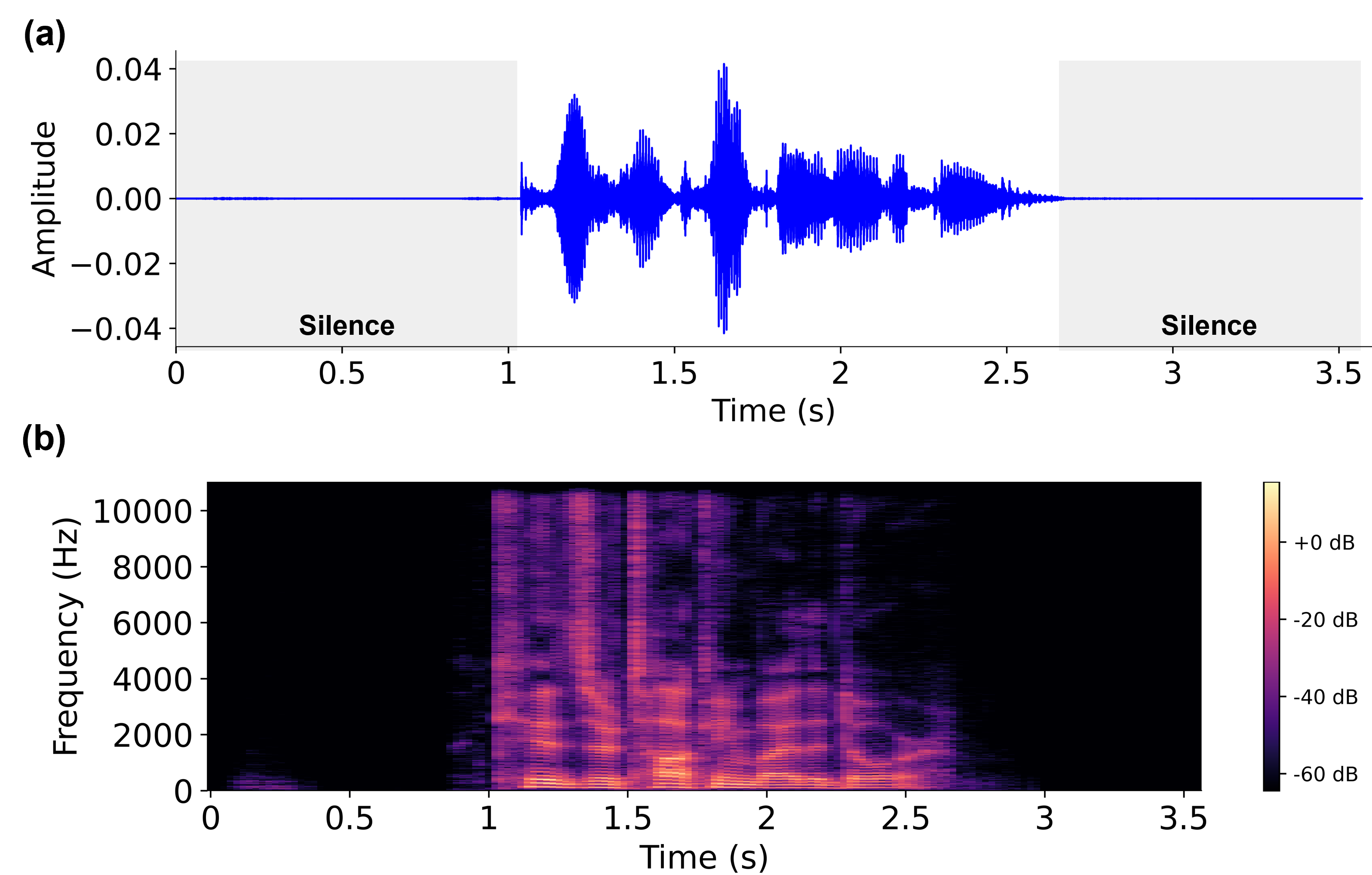}
    \caption{(a) Waveform of the audio signal showing amplitude over time, highlighting silent and active regions. (b) Spectrogram showing the frequency content over time, with color intensity indicating amplitude in decibels.}
    \label{fig:pre-processing}
\end{figure}

The Speech Emotion Recognition (SER) dataset utilized in this study consists of four publicly available benchmarks: the Crowd-sourced Emotional Multimodal Actors Dataset (CREMA-D) \cite{cao2014crema}, the Ryerson Audio-Visual Database of Emotional Speech and Song (RAVDESS) \cite{livingstone2018ryerson}, the Surrey Audio-Visual Expressed Emotion (SAVEE) \cite{jackson2014surrey}, and the Toronto Emotional Speech Set (TESS) \cite{TESS}. These datasets provide a diverse set of speech samples with annotated emotional states, including emotions such as Happy, Fear, Angry, Disgust, Surprise, Sad, and Neutral. For this study, we specifically focused on the ``Surprise" and ``Sad" emotions due to their contrasting emotional valence and the complexity they add to classification tasks (shown in Fig. \ref{fig:sad-sup-dist}). The comprehensive nature of these datasets allows for robust training and evaluation of SER classifiers, enabling the detection of emotions from vocal characteristics such as tone and pitch. This capability is critical in various applications, from customer service analysis in call centers to enhancing driver safety in automotive systems, where understanding emotional context can significantly impact performance and user experience.

\subsection{Data Pre-processing}
In this study, the audio data from the aforementioned datasets was preprocessed to enhance the performance of the SER classifier. Key features were extracted, including Zero Crossing Rate (ZCR), Root Mean Square Energy (RMS), and Mel Frequency Cepstral Coefficients (MFCC), which are instrumental in capturing the emotional nuances in speech\cite{wani2021comprehensive}. To further improve model robustness, data augmentation techniques such as noise injection, stretching, shifting, and pitching were employed to increase the variability of the training data. Each audio sample was trimmed to 2.5 seconds with a 0.6-second offset to exclude initial silent segments and focus on the relevant emotional content. The processed data was divided into training and testing sets to ensure a fair comparison. This methodological approach supports a comprehensive assessment of speech emotion detection, leveraging advanced feature extraction and augmentation techniques to enhance recognition accuracy across diverse emotional speech data.

\subsection{Classification Result and Analysis}

First, we conducted a preliminary study comparing QSVM with traditional kernel-based SVM models for speech emotion recognition of ``Surprise" and ``Sad" emotions, as shown in Table~\ref{tab:svm_comparison}. The Linear SVM, serving as a baseline, showed the lowest performance with training and testing accuracies of 55.62\% and 53.76\%, respectively, highlighting its limitations in capturing complex data patterns. Incorporating a Gaussian kernel with \( C = 1 \), the SVM improved significantly, achieving 76.88\% training and 73.75\% testing accuracy. Further optimization with \( C = 1000 \) balanced the training and testing accuracy at 84.00\%, demonstrating the impact of regularization tuning. The Central QSVM models outperformed these SVMs, with the QSVM (\( C = 1 \)) achieving 81.00\% training and 76.25\% testing accuracy. The best performance was observed with the optimized QSVM (\( C = 1000 \)), which reached 86.33\% on training and 84.33\% on testing, showcasing QSVM's superior ability to model complex data structures and outperform classical SVMs.

\begin{table}[h]
\centering
\caption{Comparison of SVM and QSVM Performance on Speech Emotion Recognition for Surprise and Sad Emotions.}
\small 
\resizebox{\columnwidth}{!}{%
\begin{tabular}{lcc}
\toprule
\textbf{Method}                              & \textbf{Train Accuracy} & \textbf{Test Accuracy} \\ 
\midrule
Linear SVM                                   & 55.62\%                 & 53.76\%                \\ 
Gaussian SVM (C = 1)                         & 76.88\%                 & 73.75\%                \\ 
Gaussian SVM (C = 1000)           & 84.00\%                 & 84.00\%                \\ 
Central QSVM (C = 1)                         & 81.00\%                 & 76.25\%                \\ 
Central QSVM (C = 1000)           & \textbf{86.33\%}        & \textbf{84.33\%}       \\ 
\bottomrule
\end{tabular}%
}
\label{tab:svm_comparison}
\end{table}

Table~\ref{tab:distributed_node_performance} presents the performance of our proposed CDQKL algorithm on the speech recognition task, demonstrating its effectiveness in enhancing classification accuracy across distributed nodes through consensus-based training. Initially, the nodes exhibited varied performance, with accuracies ranging from 62.50\% to 87.50\%. After CDQKL implementation, notable improvements were seen, particularly in local and whole test metrics. For example, Node 1's local test accuracy increased from 62.50\% to 67.50\%, and whole test accuracy improved from 71.88\% to 79.38\%. Node 4 showed the most significant gains, with local and whole test accuracies rising to 80.00\% and 78.75\%, respectively. These results illustrate CDQKL’s ability to leverage global data insights without direct data sharing, thus preserving data privacy while achieving robust performance. Compared to centralized methods, CDQKL offers similar or superior accuracy with the added advantages of distributed learning, parallel processing, and faster convergence. This aligns with prior findings on artificial and real-world datasets, positioning CDQKL as a scalable, privacy-preserving solution for distributed speech recognition tasks in quantum kernel-based machine learning.

\begin{table}[ht]
\centering
\caption{Distributed-Node Performance Metrics Before and After Training (3000 iterations \& C = 1)}
\small 
\setlength{\tabcolsep}{4pt} 
\begin{tabular}{lccc}
\toprule
\textbf{Node}      & \textbf{Metric}     & \textbf{Before Training} & \textbf{After Training}   \\ 
\midrule
\multirow{4}{*}{CDQKL Node 1} 
                   & Local Train         & 75.00\%                  & 82.50\%                   \\ 
                   & Local Test          & 62.50\%                  & 67.50\%                   \\ 
                   & Whole Train         & 77.50\%                  & 80.63\%                   \\ 
                   & Whole Test          & 71.88\%                  & 79.38\%                  \\ 
\midrule
\multirow{4}{*}{CDQKL Node 2} 
                   & Local Train         & 87.50\%                  & 92.50\%                   \\ 
                   & Local Test          & 57.50\%                  & 60.00\%                   \\ 
                   & Whole Train         & 76.25\%                  & 81.25\%                   \\ 
                   & Whole Test          & 73.13\%                  & 78.75\%                  \\ 
\midrule
\multirow{4}{*}{CDQKL Node 3} 
                   & Local Train         & 82.50\%                  & 82.50\%                   \\ 
                   & Local Test          & 77.50\%                  & 77.50\%                   \\ 
                   & Whole Train         & 75.63\%                  & 81.25\%                   \\ 
                   & Whole Test          & 72.50\%                  & 78.75\%                   \\ 
\midrule
\multirow{4}{*}{CDQKL Node 4} 
                   & Local Train         & 62.50\%                  & 82.50\%                   \\ 
                   & Local Test          & 62.50\%                  & 80.00\%                   \\ 
                   & Whole Train         & 76.83\%                  & 81.25\%                   \\ 
                   & Whole Test          & 72.50\%                  & 78.75\%                   \\ 
\bottomrule
\end{tabular}
\label{tab:distributed_node_performance}
\end{table}

\section{Discussion and conclusion}
In this work, we proposed the CDQKL approach, which enhances classification accuracy and discriminative performance across distributed nodes by exchanging model parameters between adjacent nodes without sharing local training data. Our experiments on speech recognition tasks and comparisons on artificial and real-world datasets demonstrate that CDQKL achieves comparable or superior accuracy to centralized methods while preserving data privacy, scalability, and parallel processing benefits. Despite current challenges such as device noise in quantum machine learning, our results indicate that CDQKL effectively leverages consensus-based training to incorporate global data insights, offering a promising solution for future distributed quantum computing scenarios. Future work will focus on integrating noise-aware vulnerability detection protocols, such as noise-aware detectable Byzantine agreement \cite{prest2023quantum} and error mitigation strategy \cite{chen2023short}, to enhance the robustness of CDQKL on real quantum hardware and exploring its potential within federated quantum machine learning frameworks\cite{chen2021federated}. Additionally, we aim to expand the application of CDQKL, further validating its deployment in practical quantum HPC and distributed quantum computing.


\bibliographystyle{ieeetr}
\bibliography{references}

\end{document}